# The anapole moments in disk-form MS-wave ferrite particles

E.O. Kamenetskii [*]

*Department of Electrical and Computer Engineering,
Ben Gurion University of the Negev, Beer Sheva, 84105, ISRAEL*

**Abstract**. – The anapole moments describe the parity-violating parity-odd, time-reversal-even couplings between elementary particles and the electromagnetic (EM) field. Surprisingly, the anapole-like moment properties can be found in certain artificially engineered physical systems. In microwaves, ferrite resonators with multi-resonance magnetostatic-wave (MS-wave) oscillations may have sizes two-four orders less than the free-space EM wavelength at the same frequency. MS-wave oscillations in a ferrite sample occupy a special place between the "pure" electromagnetic and spin-wave (exchange) processes. The energy density of MS-wave oscillations is not the electromagnetic-wave density of energy and not the exchange energy density as well. These "microscopic" oscillating objects – the particles – may interact with the external EM fields by a very specific way, forbidden for the classical description. To describe such interactions, the quantum mechanical analysis should be used. The presence of surface magnetic currents is one of the features of MS oscillations in a normally magnetized ferrite disk resonator. Because of such magnetic currents, MS oscillations in ferrite disk resonators become parity violating. The parity-violating couplings between disk-form ferrite particles and the external EM field should be analyzed based on the notion of an anapole moment.

---

[*] e-mail address: kmntsk@ee.bgu.ac.il

*Introduction*. – The fundamental discrete symmetries of parity (P), time reversal (T) and charge conjugation (C), and their violations in certain situations, are central in modern elementary particle physics, and in atomic and molecular physics. As a basic principle, the weak interaction is considered as the only fundamental interaction, which does not respect left-right symmetry. Atoms are chiral due to the parity-violating weak neutral current interaction between the nucleus and the electrons [1]. As an example of chirality (helicity) in atomic phenomena, one can consider the anapole-moment properties. The anapole moment is spin dependent. It takes place in systems with the parity violation and with the annual magnetic field [2]. The anapole moment plays the essential role in nuclear helimagnetism [3,4]. It was considered as an intrinsic property of a diatomic polar molecule [5].

The problems of fundamental discrete symmetries constitute, in particular, the subject of strong discussions in the field of the magnetoelectric (ME) effect and chiral (optically active) media. Recently, the notion of the anapole was used for explanation of the ME effect in solid-state magnetic crystals [6]. The nature of the atomic chirality, however, is not the same as the chirality of enantiomers. Chiral molecules (being the delocalized structures) have a well-defined form and so they must be in one of the configurations. It has been made many theoretical predictions that enantiomers of chiral molecules have different spectra because of parity violation associated with neutral currents in the weak interaction [7] or with the particle-antiparticle replacement in different enantiomers [8]. Many spectroscopic proposals have been made to observe a parity violation effect in molecules (see e.g. [9]). However, considering the weakness of the effects, it is broadly admitted that the weak interaction may be neglected in molecular physics and even more clearly in chemistry and biology. Moreover, in a quantum mechanical description of optical activity, the remaining questions are: "Can the existence of chiral molecules be explained in quantum mechanics? Why the chiral molecules are non-stable and never found in energy eigenstates?" [10,11].

It becomes clear now that the P, T, C symmetries should be also one of the central questions in physics of novel artificial materials, in particular, chiral and bianisotropic composite materials. In consideration of physical properties of structural elements of such materials – artificial atoms – one should use the quantum-mechanical-like description [12]. At present, there is a strong interest in different quantum phenomena on a macroscopic scale. For example, "macroscopic" flux quantization, "macroscopic" quantum tunneling and resonance tunneling, macroscopic quantum effects in Josephson-junction circuits – all these effects constitute a subject of extensive and intensive fundamental research and also a subject of conceivable applications of quantum-state engineering (see e.g. [13] and references therein). Another aspects of "macroscopic" quantum phenomena – a probability amplitude theory in the classical macrodomain of the dynamics of charged particles in a magnetic field – were considered in [14].

In microwaves, ferrite resonators with multi-resonance MS-wave oscillations may have sizes two-four orders less than the free-space EM wavelength at the same frequency [15]. These "microscopic" oscillating objects – the particles – may interact with the external EM fields by a very specific way, forbidden for the classical description. To describe such interactions, the quantum mechanical analysis should be used. One of examples of the quantum-mechanical-like effects in small ferrite resonators is the ME effect. Recent experimental results demonstrate the ME effect in normally magnetized MS-wave flat ferrite resonators with surface electrodes [16-19]. Among different types of ferrite ME particles examined in [16-19] one should, certainly, distinguish the structures based on ferrite disks with linear (one-dimensional) surface electrodes. In this case, the spectrums excited by different types of RF fields (the $\vec{E}, \vec{H}$ fields or their combinations) have the same positions of the main oscillation peaks. Because of these spectral properties one can talk about the *unified process of ME oscillations*. This fact makes possible to characterize a disk + wire - type ferrite ME particle by a set of parameters and represent it as a "point source" with the locally coupled two (electric and magnetic) dipole moments. These ferrite particles provide the P, T, C symmetry characteristics distinguished from those produced by point sources of the EM fields [20,21].

It becomes clear that the main physics of the effective *quasielectrostatic to quasimagnetostatic energy transformation* (and vice versa) in ferrite ME particles and the symmetry properties of ME fields originated by these particles, observed in [16-19], should be found from the spectral properties of MS oscillations in normally magnetized disk ferrite resonators. So, initially, we have to focus our attention to a "pure" (without a metal electrode) ferrite resonator. MS oscillations in a small ferrite disk resonator could be characterized by a discrete spectrum of energy levels [22]. Calculations of the energy spectra of MS oscillations in ferrite disks were made in [23]. In paper [24] we consider surface magnetic currents, which characterize oscillations MS in a normally magnetized ferrite disk and give parity-violating perturbations. As we discuss in the present paper, because of such magnetic currents, one has the parity-odd, time-reversal-even motion processes having a clear analogy with the anapole-moment characteristics in the weak interaction.

Recent experiments show that MS oscillations in a normally magnetized ferrite disk are strongly affected by a normal component of the external RF electric field [25]. The observed multi-resonance process cannot be characterized as the induced electric polarization effect in a particle. There are the *eigen-electric-moment* oscillations caused by the motion processes in a ferrite resonator. Since the RF electric field does not change sign under time inversion, the eigen electric moment should also be characterized by the time-reversal-even properties. This is the case of an anapole moment, considered, for the first time, by Zel'dovich [2].

*Circular surface magnetic currents in MS-wave ferrite disks*. – MS-wave oscillations in a ferrite sample occupy an intermediate place between the "pure" electromagnetic and spin-wave (exchange) processes. For the characteristic specimen size $l$, MS-wave oscillations take place when

$$\sqrt{\alpha} \ll l \ll c/\omega, \qquad (1)$$

where $\sqrt{\alpha}$ is the characteristic length of the exchange energy [26]. Relation (1) means that we neglect all the electromagnetic effects (the effects connected with a finite velocity of propagation of electromagnetic perturbations) and, at the same time, all the exchange processes (the effects connected with spatial dispersion in ferrites). The energy density of MS-wave oscillations is not the electromagnetic-wave density of energy and not the exchange energy density as well. MS waves and oscillations in ferrites are described based on the Walker equation – the second-order differential equation written for MS potential $\psi$ ($\vec{H} = -\nabla \psi$). In an unbounded ferrite, MS waves at the oscillating frequencies have any wavelength. The frequency degeneracy is removed by considering the effects of finite sample boundaries. It is supposed that the boundary conditions are that the tangential component of $\vec{H}$ and the normal component of $\vec{B}$ should be continuous across the boundary of the specimen [15].

In a normally magnetized ferrite-disk resonator with a small thickness $h$ to diameter $2R$ ratio, the monochromatic MS-wave potential function $\psi$ is represented as [22]:

$$\psi = \sum_{p,q} A_{pq} \tilde{\xi}_{pq}(z) \tilde{\varphi}_q(\rho, \alpha), \qquad (2)$$

where $A_{pq}$ is a MS mode amplitude, $\tilde{\xi}_{pq}(z)$ and $\tilde{\varphi}_q(\rho, \alpha)$ are dimensionless functions describing, respectively, "thickness" (z coordinate) and "in-plane", or "flat" (radial $\rho$ and azimuth $\alpha$ coordinates) MS modes. In a ferrite disk with a small thickness to diameter ratio, the spectrum of "thickness" modes is enough "rare" compared to the "dense" spectrum of "flat" modes [17,22,23].

Based on the Walker equation, one has for every "flat" MS mode:

$$\hat{G}_\perp \tilde{\varphi}_q = \beta_q^2 \tilde{\varphi}_q, \tag{3}$$

where

$$\hat{G}_\perp \equiv \mu \nabla_\perp^2, \tag{4}$$

$\nabla_\perp^2$ is the two-dimensional, "in-plane", Laplace operator and $\beta_q$ is the propagation constant along z-axis, $\mu$ is the diagonal component of the permeability tensor. For propagating MS modes, operator $\hat{G}_\perp$ is the positive definite operator. Since a two-dimensional ("in-plane") differential operator $\hat{G}_\perp$ contains $\nabla_\perp^2$ (the two-dimensional, "in-plane", Laplace operator), a double integration by parts (the Green theorem) on $S$ – a square of an "in-plane" cross section of an open ferrite disk – of the integral $\int (\hat{G}_\perp \tilde{\varphi}) \tilde{\varphi}^* \, dS$, gives the following boundary condition for the energy orthonormality:

$$\mu \left( \frac{\partial \tilde{\varphi}}{\partial \rho} \right)_{\rho=R^-} - \left( \frac{\partial \tilde{\varphi}}{\partial \rho} \right)_{\rho=R^+} = 0 \tag{5}$$

or

$$\mu (H_\rho)_{\rho=R^-} - (H_\rho)_{\rho=R^+} = 0, \tag{6}$$

where $H_\rho = -\frac{\partial \psi}{\partial \rho}$ is a radial component of the RF magnetic field, $R^-$ and $R^+$ designate, respectively, the inner (ferrite) and outer (dielectric) regions of a disk resonator with radius R.

For operator $\hat{G}_\perp$, the boundary condition of the MS-potential continuity together with boundary condition (5) [or (6)] are the so-called essential boundary conditions [27]. When such boundary conditions are used, the MS-potential eigen functions of operator $\hat{G}_\perp$ form a *complete basis in an energy functional space*, and the functional describing an average quantity of energy, has a minimum at the energy eigenfunctions [27]. The essential boundary conditions differ from the homogeneous electrodynamics boundary conditions at $\rho = R$, which demand continuity for the radial component of the magnetic flux density (together with continuity for potential $\widetilde{\varphi}$). The last ones are called as natural boundary conditions [27]. In [23], calculations of complete-set energy spectra of MS oscillations in a ferrite disk resonator were made based on the essential boundary conditions.

In a cylindrical coordinate system, continuity for a radial component of the magnetic flux density (the natural boundary condition) at $\rho = R$ is described as

$$\mu(H_\rho)_{\rho=R^-} - (H_\rho)_{\rho=R^+} = -i\mu_a (H_\alpha)_{\rho=R^-} \qquad (7)$$

where $H_\alpha = -\frac{1}{\rho}\frac{\partial \psi}{\partial \alpha}$ is an azimuth component of the RF magnetic field, $\mu_a$ is the off-diagonal component of the permeability tensor. Supposing that $\widetilde{\varphi} \sim e^{-i\nu\alpha}$, one can rewrite (7) as

$$\mu\left(\frac{\partial \widetilde{\varphi}}{\partial \rho}\right)_{\rho=R^-} - \left(\frac{\partial \widetilde{\varphi}}{\partial \rho}\right)_{\rho=R^+} = -\frac{\mu_a}{R}\nu(\widetilde{\varphi})_{\rho=R^-} \quad . \qquad (8)$$

One can see that "in-plane" functions $\widetilde{\varphi}$, being determined by two second-order differential equations (the Bessel equations for functions $\widetilde{\varphi}$ inside and outside the ferrite) and one first-order differential equation (8), are dependent on both a quantity and a sign of $\nu$. So the functions $\widetilde{\varphi}$ *cannot be the single-valued functions* for angle $\alpha$ varying from 0 to $2\pi$. In other words, we have different results for positive and negative directions of an angle coordinate when $0 \leq \alpha \leq 2\pi$.

Let us consider a circulation of vector $\vec{H}$ along a circular contour $L = 2\pi R$. Since on the contour L, $H_\alpha^{(L)} = -\frac{1}{R}\frac{\partial \widetilde{\varphi}^{(L)}}{\partial \alpha}$, we can write this circulation as $C = \nu \int_0^{2\pi} \widetilde{\varphi}^{(L)} d\alpha$. The circulation C should be equal to zero for a single-valued function $\widetilde{\varphi}$. This fact also follows from the MS description ($\nabla \times \vec{H} = 0$). Our analysis shows, however, that this circulation has a non-zero quantity. The solution depends on both a modulus and a sign of $\nu$. We have a sequence of angular eigenvalues restricted from above and below by values equal in a modulus and different in a sign, which we denote as $\pm s^e$. The difference $2s^e$ between the largest and smallest values is an integer or zero. So $s^e$ can have values $0, \pm 1/2, \pm 1, \pm 3/2, ..., \nu/2$. At a full-angle "in-plane" rotation (at an angle equal to $2\pi$) of a system of coordinates, the "in-plane" functions $\tilde{\varphi}$ with integer values $s^e$ return to their initial states (single-valued functions) and "in-plane" functions $\tilde{\varphi}$ with the half-integer values $s^e$ will have an opposite sign (double-valued functions). The only possibility in our case is to suggest that $s^e$ are the half-integer quantities. Because of the double-valuedness properties of MS-potential functions on a lateral surface of a ferrite disk resonator, we can talk about the "spinning-type rotation" along a border contour L [24]. Along with the well-known notion of the "magnetic spin" as a quantity correlated with the eigen magnetic moment of a particle, we introduced in [24] the notion of the "electric spin" as a quantity correlated with the eigen electric moment. For integer quantities $s^e$ the eigen electric moment is equal to zero, but it is non-zero for half-integer values $s^e$. From the above consideration it becomes clear that superscript e in $s^e$ means "electric".

The main feature of boundary condition (7) arises from the quantity of an azimuth magnetic field in the right-hand side. One can see that this is a singular field, which exists only in an infinitesimally narrow cylindrical layer abutting (from a ferrite side), to the ferrite-dielectric border. One does not have any special conditions connecting radial and azimuth components of magnetic fields on other (inner or outer) circular contours, except contour L. Because of the annual magnetic field, arising from boundary condition (7), the notion of an effective circular magnetic current can be considered.

Let us formally introduce a quantity of a magnetic current:

$$\vec{j}^m(z) \equiv \frac{1}{4\pi} i\omega\mu_a \vec{H}_\alpha(z) \qquad (9)$$

We can rewrite the boundary condition (7) as follows:

$$\delta(\rho - R)\left[\frac{1}{4\pi}\omega\mu(H_\rho)_{\rho=R^-} - \frac{1}{4\pi}\omega(H_\rho)_{\rho=R^+}\right] = -i^m \qquad (10)$$

where $i^m$ is a density of an effective boundary magnetic current defined as

$$i^m(z) \equiv \delta(\rho - R)\frac{1}{4\pi} i\omega\mu_a (\vec{H}_\alpha(z))_{\rho=R^-} = \delta(\rho - R)\vec{j}^m(z) \qquad (11)$$

In a supposition that "flat" functions $\tilde{\varphi}$ form a complete basis in the energy functional space with use of boundary condition (5) [or (6)], it becomes evident that the effective boundary magnetic current slips from the main properties of this functional space. This current cannot be considered as a single-valued function. In the description of the MS-potential functions in a ferrite disk, taking into account the effective surface magnetic current, certain additional coordinates should be used. It means that additional eigenvalues and eigenfunctions should appear on boundary contour L. One can see that for modes with zero azimuth variations ($\nu = 0$), both the essential and natural boundary conditions are the same, being described by Eqn. (5) [or (6)]. In this case, the effective boundary magnetic current is equal to zero.

The matrices characterizing the components of the "electric spin" we represent as:

$$\hat{s}^e_+ = \begin{pmatrix} 0 & 0 \\ 1 & 0 \end{pmatrix}, \quad \hat{s}^e_- = \begin{pmatrix} 0 & 1 \\ 0 & 0 \end{pmatrix}, \quad \hat{s}^e_z = |w|\begin{pmatrix} 1 & 0 \\ 0 & -1 \end{pmatrix}. \qquad (12)$$

Quantity w characterizes the "spin coordinates". To distinguish the "right" and "left" MS-potential functions at a full-angle "in-plane" rotation (at an angle equal to $2\pi$) of a system of coordinates, we should write that $w = k\frac{1}{2}$, where k is an integer odd (positive or negative) quantity. One should suppose that to have the system stability, the "spin states", characterizing by quantities w, ought to be in a certain synchronism with the "orbit states", characterizing by quantities $\nu$.

We consider now the "border" MS-potential "flat" functions $\widetilde{\widetilde{\varphi}}$ on contour L. There are singular functions describing the "spin states". For k-th "border" eigenfunction we can write

$$\widetilde{\widetilde{\varphi}}_k = B_k e^{-i w_k \alpha}, \tag{13}$$

where $B_k$ is an amplitude coefficient. Introducing function $\widetilde{\widetilde{\varphi}}$, we have to note that this is not an independent function with respect to function $\widetilde{\varphi}$, but the function showing certain additional properties, additional states of the MS-potential scalar wave function. For a certain "thickness" mode and a certain "flat" mode we can represent the $\alpha$-component of the "border" (singular) magnetic field as

$$\left(H_\alpha(z)\right)_{\rho=R^-} = -A\widetilde{\xi}(z)\nabla_\alpha \widetilde{\widetilde{\varphi}} = -A\widetilde{\xi}(z)\frac{1}{R}\frac{\partial \widetilde{\widetilde{\varphi}}}{\partial \alpha}\bigg|_{\rho=R^-}. \tag{14}$$

For a circular effective boundary magnetic current we have now [see Expr. (11)]:

$$\left(i^m(z)\right)_k = -A\widetilde{\xi}(z)\frac{i\omega\mu_a}{4\pi R}\frac{\partial \widetilde{\widetilde{\varphi}}}{\partial \alpha}\bigg|_{\rho=R^-} = -A\widetilde{\xi}(z)\frac{\omega\mu_a}{4\pi R}w_k B_k e^{-i w_k \alpha} \tag{15}$$

The circular surface magnetic current does not exist due to only precession of magnetization. It appears because of the combined effect of precession in a ferrite material and "spinning rotation" caused by the special-type boundary conditions. The "border" MS-potential functions $\widetilde{\widetilde{\varphi}}$, being characterized by the "spin coordinates", are antisymmetrical functions. At the same time, as it follows from Expr. (15), the effective magnetic currents are described by symmetrical functions with respect to the "spin coordinates". In other words, the effective magnetic current has the same direction for the "right" and "left" spinning states. The signs of magnetic current $i^m$ are different for different signs of $\mu_a$. However, the "positive" ($\mu_a > 0$) and "negative" ($\mu_a < 0$) magnetic currents do not mutually compensate each other since for different signs of $\mu_a$ we have structures with different symmetries.

Circulation of current $i^m$ along contour L gives a nonzero quantity when $w_k$ is a number divisible by $\frac{1}{2}$:

$$D_k(z) = \oint_L (i^m)_k \, dl = R \int_0^{2\pi} (i^m)_k \, d\alpha = iA\tilde{\xi}(z)\frac{\omega \mu_a}{2\pi} B_k. \tag{16}$$

Since circulation $D_k(z)$ is a non-zero quantity, one can define an electric moment of a whole ferrite disk resonator (in a region far away from a disk) as

$$a_k^e = -i\frac{1}{2c}\int_0^h dz \oint_{L=2\pi R} (\vec{\rho} \times \vec{i}^m) \cdot \vec{e}_z \, dl = A\frac{\omega \mu_a}{4\pi c} RB_k \int_0^h \tilde{\xi}(z)dz, \tag{17}$$

The off-diagonal component of the permeability tensor, $\mu_a$, can be correlated with a magnetic vector of gyration [15]:

$$\vec{g}^m = \frac{\mu_a}{4\pi}\vec{e}_z, \tag{18}$$

where $\vec{e}_z$ is the unit vector along z-axis. A sign of $g^m$ corresponds to a sign of $\mu_a$. A sign of amplitude $B$ depends on orientation of vector $\vec{s}^e$ ($\vec{s}^e = s^e \vec{e}_z$) with respect to z-axis. So one can distinguish two cases: $\vec{s}^e \cdot \vec{g}^m > 0$ and $\vec{s}^e \cdot \vec{g}^m < 0$.

The property of helicity is well-known in elementary particle physics. There is no left-right symmetry since spin orientation is not separated from orientation of a linear momentum. Following discussions in paper [2], one can see that in such a case a particle cannot be affected by the electric field. In our case, the spin orientation $\vec{s}^e$ is not separated from orientation of a linear momentum $\vec{a}^e$, but taking into account also orientation of vector $\vec{g}^m$. Figs. 1 (a) and (b) show two possible situations for cases $\vec{s}^e \cdot \vec{g}^m > 0$ and $\vec{s}^e \cdot \vec{g}^m < 0$. In both cases, a sign of an electric moment $a^e$ is invariant with respect to the time reversal (both vectors $\vec{s}^e$ and $\vec{g}^m$ are the axial vectors). This fact is illustrated in Figs. 1 (c) and (d). Our analysis gives an evidence for the anapole-moment properties of vector $\vec{a}^e$. It is necessary to note that together with the pictures of two-vector (vectors $\vec{s}^e$ and $\vec{g}^m$) orientations with respect to orientation of a bias magnetic field $\vec{H}_0$ shown in Figs. 1 (a), (b), another two pictures can be demonstrated. So, in a general case, the analysis should be expanded with consideration of four possible cases.

Following Zel'dovich [2], the so-called toroidal (or anapole) moment is an odd parity magnetic field distribution of rank 1 (dipole). One can trace a clear analogy between the above consideration and electromagnetic properties of a toroidal solenoid. If we picture an element of magnetic current $i^m$ as a small electric-current loop, the combination of all elements along contour L can be viewed (for a ferrite disk with a small thickness to diameter ratio) as a toroidal current winding producing an anapole moment. A similar situation, when the magnetization current is entirely responsible for the anapole moment, one can find in [3].

Now the question about a nature of a source of magnetic current $i^m$ arises. As we discussed in [24], with consideration of non-zero circulation along contour L we refer, in fact, to the concept of non-integrable, i.e. path-dependent, phase factor defined by an integral taken around an unshrincable loop. Since the difference between $\nu$ and w is a non-integer quantity, one has the electric field described by a non-single-valued function with respect to a frame of reference of the "spinning coordinates" (determining function $\tilde{\tilde{\varphi}}$). This leads to existence of an electric flux $\Phi^E$ through the "surface ring" opening. Our situation becomes resembling (and dual) to the case of a mesoscopic metal ring threaded by a magnetic flux when the flux changes with time [28]. Based on this analogy one can say that in our case, a persistent ring magnetic current is due to the magnetomotive force expressed as a Maxwell-law generator as follows:

$$\varepsilon^m(t) = \frac{d\Phi^E}{dt} \qquad (19)$$

This quantity defines amplitude $B_k$.

*Helical surface magnetic currents in MS-wave ferrite disks*. – In the above model we made use of a cylindrical coordinate system. In this case we had to introduce the notion of "spinning rotation", which arises from the necessity to utilize the double-valued MS-potential functions on a border contour L. The question, however, is still open: What is the reason of apearence of such double-valued functions? The answer can be found from the fact that we considered rotation separately from an axial motion. In other words, the above consideration was based on an assumption that surface magnetic currents in all the elementary rings (having elementary thickness dz) are in phase. In this assumption we were able to define the anapole moment by integration over z [see Expr. (17)].

Our analysis suggests an idea that in another coordinate system – the helical coordinate system – the picture could be different from the above representation. One can suppose that in a case of helical surface magnetic currents (which we have in a helical coordinate system) the anapole moment will arise not from the property of "spinning rotation", but from the combined boundary effect on contour L and planes $z = 0$ and $z = h$ of a ferrite disk. To illustrate this statement let us examine distributions of helical surface magnetic currents for possible manifestation of the anapole-moment properties.

Since helical surface magnetic current $i^m$ takes place due to a combined effect of a circular and axial motion processes, the upper ($z = h$) and lower ($z = 0$) points should be the turn points. So the current circumscribes a bifilar helix. As far as we have a distributed-parameter-system, the surface current is described by the standing-wave characteristics. It means that there are knots and maximums in the current distributions. Two alternatives can be traced in this model: (A) current $i^m$ has a maximum amplitude in the turn points, and (B) current $i^m$ is equal to zero in the upper ($z = h$) and lower ($z = 0$) points. For a possible manifestation of the anapole-moment properties, the case (A) does not represent any interest, but the case (B) does. When current $i^m$ is equal to zero in the upper ($z = h$) and lower ($z = 0$) points, it changes its sign (both with respect to azimuth and axial coordinates) in these points. It means that in the turn points the left-handed helix is transformed to the right-handed helix and vice versa.

Suppose that we have the helical magnetic current distribution corresponding to case (B). Let vector $\vec{p}$ be unit vector along z-axis, showing direction of an axial motion. Suppose that an axial motion is along vector $\vec{g}^m$. So we have $\vec{p} \cdot \vec{g}^m > 0$. Assume that for the "up" axial motion (the motion along z-axis) we have a certain direction of helical current $i^m$, which corresponds to a right-handed helix (Fig. 2 (a)). The case when vector $\vec{g}^m$ is anti-parallel to vector $\vec{p}$ ($\vec{p} \cdot \vec{g}^m < 0$) is not identical to the considered above. An example of the "up" axial motion along vector $-\vec{g}^m$ with current $i^m$ circumscribing a left-handed helix is shown in Fig. 2 (b). Now let us perform the time reversal operation. Since $\vec{g}^m$ is an axial vector, it has an opposite direction with respect to z-axis, when the time reversal operation is made. Because of correlation between directions of vectors $\vec{p}$ and $\vec{g}^m$ (due to the Onsager principle), an opposite sign of $g^m$ means the "down" axial motion (the motion along $-z$ axis). With the motion "down" we also have helical surface currents but with an opposite direction (see Figs. 2 (c) and (d)). In both cases (Fig. 2 (c) versus Fig. 2 (a) and Fig. 2 (d) versus Fig. 2 (b)) there are the same directions of anapole moments, produced by helical surface magnetic currents. So the time-reversal symmetry does not forbid the manifestation of the anapole-moment properties. It is interesting to note that following the above picture of representation of an element of magnetic current $i^m$ as a small electric-current loop, the combination of all elements along a helix can be viewed as a helical-form (not toroidal) "doughnut" current winding. One has also to take into account that together with the surface magnetic current distribution, the above model should describe the magnetic charge density distribution as well. For a case when current $i^m$ is equal to zero in the turn points, the continuity equation gives maximums of the magnetic charge density in these points.

In a helical coordinate system the azimuth number does not necessarily have to be an integer as it would in a standard cylindrical system [29]. One can suppose that in our case of a flat ferrite disk the azimuth number might be as a half integer quantity. So the model of a helical magnetic current can be properly reduced to the model of a circular magnetic current with a clear explanation why the "spinning rotation" on boundary contour L takes place. To find a helical surface magnetic current, one has to rewrite the boundary conditions (7) and (8) in the so-called helical coordinate system [29]. Unlike the cylindrical coordinate system, in the helical system, two different types of solutions are admitted, one right-handed and one left-handed. To express properly the helical surface magnetic current, it is necessary to represent in a helical coordinate system differential equations for the MS-potential function: the Walker equation inside a ferrite and the Laplace equation outside a ferrite. Such an analysis is beyond the frames of this paper and is the subject of our future investigations.

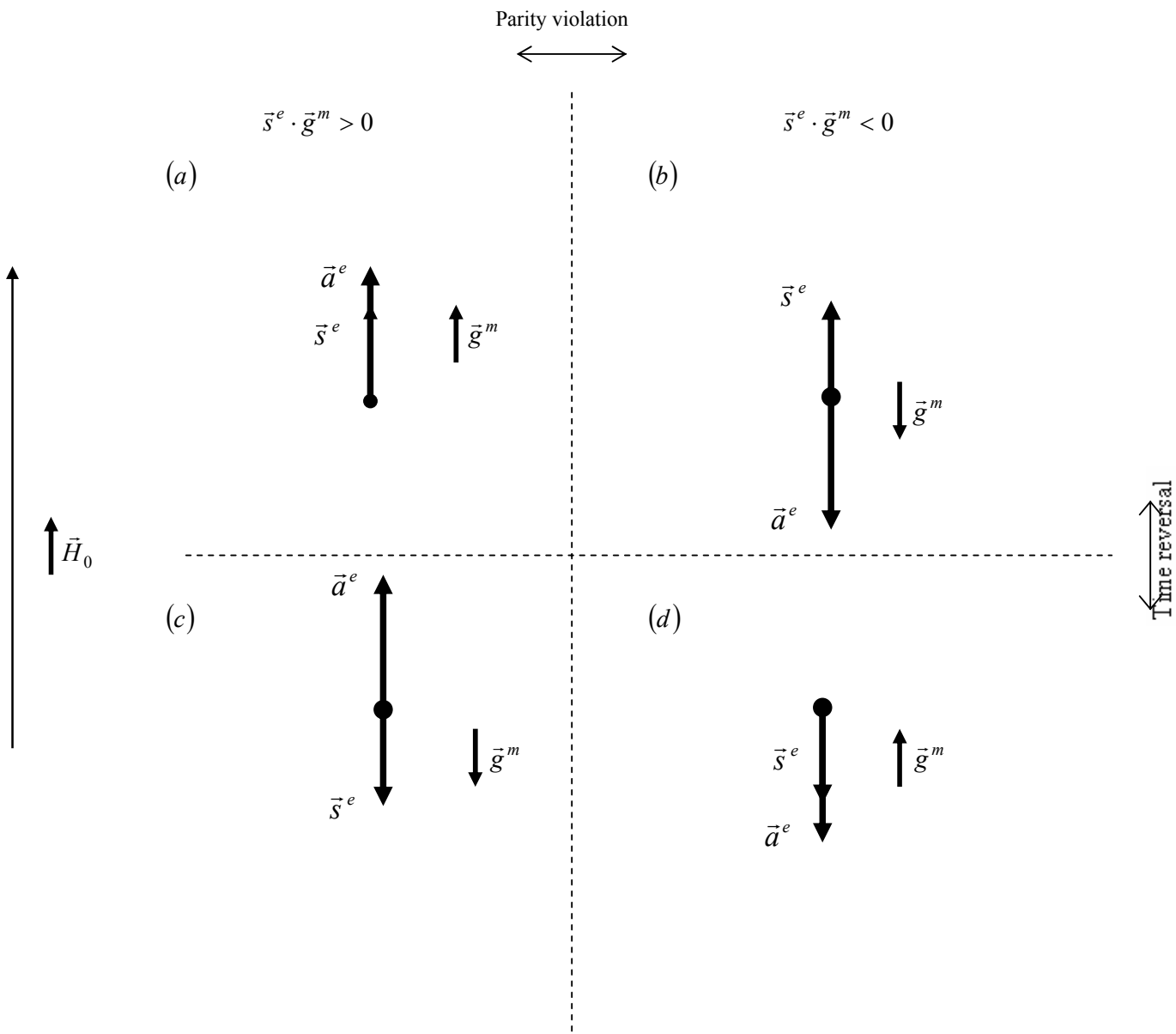

Fig. 1

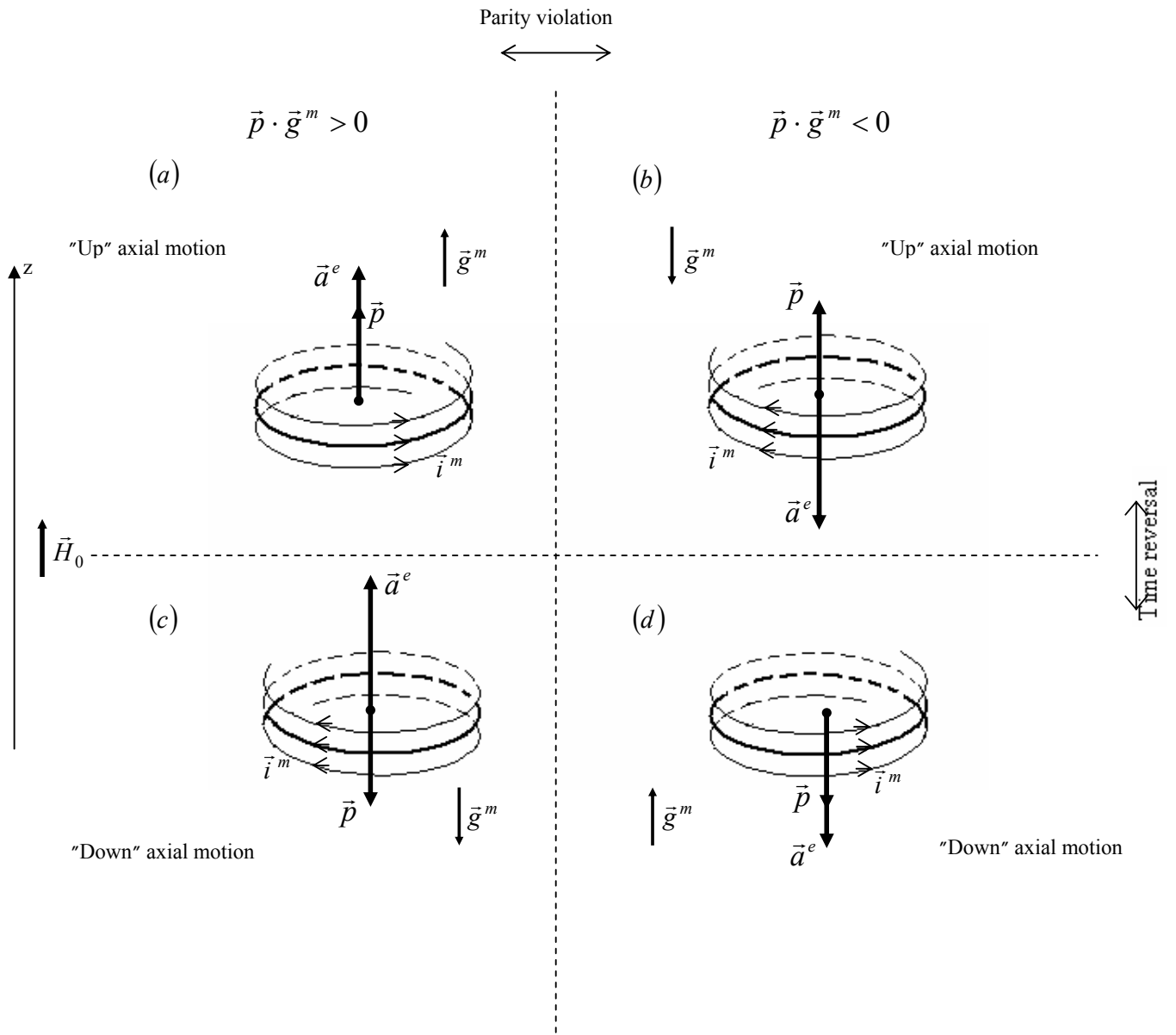

Fig. 2

The anapole moments in disk-form MS-wave ferrite particles

by E.O. Kamenetskii

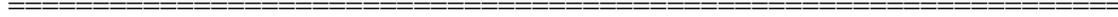

Figure captions

Fig. 1 The anapole-moment properties in MS-wave ferrite disk resonator for circular surface magnetic currents.

Fig. 2 The anapole-moment properties in MS-wave ferrite disk resonator for helicoidal surface magnetic currents.

To show distribution of the magnetic current amplitude along z-axis, the current helices are drawn by lines with different widths. Boldface lines correspond to maximum current amplitudes.